\documentclass[11pt,oneside]{article}
\usepackage{graphicx}
\begin{document}
 
\title{{\bf FORMATION OF GALACTIC SYSTEMS IN LIGHT OF THE MAGNESIUM ABUNDANCE
IN FIELD STARS: THE THIN DISK}}
\author{{\bf V.~A.~Marsakov, T.~V.~Borkova}\\
Institute of Physics, Rostov State University,\\
194, Stachki street, Rostov-on-Don, Russia, 344090\\
e-mail: marsakov@ip.rsu.ru, borkova@ip.rsu.ru}
\date{accepted \ 2006, Astronomy Letters, Vol. 32 No. 6, P.376-392}
\maketitle

\begin {abstract}
Data from our compiled catalog of spectroscopically determined 
magnesium abundances in stars with accurate parallaxes are used to 
select thin-disk dwarfs and subgiants according to kinematic criteria. 
We analyze the relations between the relative magnesium abundances in 
stars, [Mg/Fe], and their metallicities, Galactic orbital elements, and 
ages.The [Mg/Fe] ratios in the thin disk at any metallicity in the range 
($-1.0<[Fe/H ]<-0.4$~dex) are shown to be smaller than those in 
the thick disk implying that the thin-disk stars are on average 
younger than the thick-disk stars. The relative magnesium abundances
in such metal-poor thin-disk stars have been found to systematically 
decrease with increasing stellar orbital radii in such a way that 
magnesium overabundances ($[Mg/Fe ]>0.2$~dex) are essentially observed 
only in the stars whose orbits lie almost entirely within the solar circle.
At the same time, the range of metallicities in magnesium-poor stars 
is displaced from ($-0.5<[Fe/H ]<+0.3$~dex) to ($-0.7<[Fe/H ]<+0.2$~dex)
as their orbital radii increase. This behavior suggests that, first, the 
star formation rate decreases with increasing Galactocentric distance and,
second, there was no star formation for some time outside the solar
circle while this process was continuous within the solar circle. The 
decrease in the star formation rate with increasing Galactocentric 
distance is responsible for the existence of a negative radial 
metallicity gradient (grad$_{R}$[Fe/H]=($-0.05 \pm 0.01$)~kpc$^{-1}$) in 
the disk, which shows a tendency to increase with decreasing
age. At the same time the relative magnesium abundance exhibits no 
radial gradient. We have confirmed the existence of a steep negative 
vertical metallicity gradient (grad$_Z$[Fe/H]=($-0.29 \pm 0.06$)~kpc$^{-1}$)
and detected a significant positive vertical gradient in relative 
magnesium abundance (grad$_Z$[Mg/Fe]=($0.13 \pm 0.02$)~kpc$^{-1}$); both 
gradients increase appreciably in absolute value with decreasing age. We 
have found that there is not only an age--metallicity relation, but also 
an age--magnesium abundance relation in the thin disk. We surmise that 
the thin disk has a multicomponent structure but the existence of a
negative trend in the star formation rate along the Galactocentric 
radius does not allow the stars of its various components to be identified 
in the immediate solar neighborhood.
\\


\end {abstract}

\section*{Introduction}

This paper is a continuation of our systematic description of 
chemical and spatial--kinematic properties of the stars that 
are currently in the solar neighborhood,but belong to different 
Galactic subsystems, based on data from our compiled catalog of 
spectroscopically determined elemental abundances in stars
with accurate parallaxes (Borkova and Marsakov~2005). Previously 
(Marsakov and Borkova~2005), we analyzed the properties of 
thick-disk stars. 

The relative atmospheric elemental abundances of low-mass 
main-sequence stars can be used to estimate the parameters of the 
initial mass function and the star formation rate 
and as the time scale of a chemically evolving system. Thus, for 
example, the $\alpha$-elements (O, Mg, Si, S, Ca and Ti) together 
with a small number of iron atoms are currently believed to be 
synthesized in the high-mass ($M>10M_\odot$) asymptotic-giant-branch 
(AGB) progenitors of type\,II supernovae (Arnett~1978), while the 
bulk of the iron-group elements are produced during type\,Ia 
supernova explosions through mass accretion onto a carbon--oxygen white 
dwarf in a close binary system (Thielemann et al.~1986). The 
theoretically predicted yield of $\alpha$-elements in SNe\,II 
increases with presupernova mass (Woosley and Weaver~1995); therefore,
the larger the shift of the initial mass function toward the higher 
masses, the larger the observed [$\alpha$/Fe] ratio in the atmospheres 
of the metal-poorest stars. Beginning from the paper by Tinsley~(1979),
the negative trend in the [$\alpha$/Fe] ratio as a function of metallicity
observed in the Galaxy has been assumed to be due to a difference in 
the evolution times of these stars. Indeed,the evolution time scale for 
type\,II presupernovae is only $\approx 30$~Myr while massive SNe\,Ia
explosions begin only in $\approx (0.5\div 1.5$)~Gyr (Matteucci and 
Greggio~1986; Yoshii et al.~1996). The higher the star formation rate in 
the system, the larger the metallicity at which the knee attributable 
to the onset of SNe\,Ia explosions which result in an enrichment
of the interstellar medium with iron-group elements, will be observed in 
the [$\alpha$/Fe]--[Fe/H] relation. The lower the star formation rate in 
the system, the steeper the further decrease in the [$\alpha$/Fe] ratio 
with increasing total metallicity. If star formation in the system is
halted altogether then the source of $\alpha$-elements (i.\,e., SNe\,II)
will vanish and only SNe\,Ia will enrich the interstellar medium with 
iron-group elements; therefore the [$\alpha$/Fe] ratio will decrease 
suddenly (Wyse and Gilmore~1991).

Since more than 90\,\% of the stars in the immediate solar neighborhood belong 
to the youngest (in the Galaxy) thin-disk subsystem (Buser et al.~1999),
the chemical composition of this subsystem has been studied in greatest detail.
However the sizes of the original samples in all works were very limited;
this is probably the reason why the results and conclusions are occasionally 
in conflict with one another. The work of Edvardsson et al.~(1993) is a
classical analysis of a sample of nearby stars containing $\approx 200$~F--G 
dwarfs with metallicities [Fe/H]$>-1.0$. In particular this work revealed a 
decrease in the [$\alpha$/Fe] ratio with increasing mean Galactocentric
distance $R_m$ for stars in the range $0.8 <[Fe/H ]<0.4$ which was explained 
by a decrease in the star formation rate from the center to the periphery of the
Galactic disk. Nissen~(2004) disputes this explanation and argues that this 
trend is attributable solely to the fundamental difference between the relative
abundances of $\alpha$-elements in the thin-disk and thick-disk stars.
Indeed, Edvardsson et al.~(1993) did not separate their stars into the two 
subsystems, while Fuhrmann et al.~(1995), Fuhrmann~(1998), Nissen and Shuster~(1997),
and Gratton et al.~(2000) showed that the relative abundances of $\alpha$-elements 
decrease abruptly during the transition from the thick-disk to thin-disk stars.
This indicates that star formation is delayed, i.\,e. the transition between the 
disk subsystems is discrete. Other studies show that there are stars in the 
thick disk with the same low relative abundances of $\alpha$-elements as those 
in the thin disk; conversely there are stars in the thin disk with the
same high relative abundances as those in the thick disk. Some of the authors 
(see, e.\,g. Feldzing et al.~2003; Bensby et al.~2004) argue that the [$\alpha$/Fe]
ratio in the thin disk at the same metallicity (up to its solar value)is 
systematically lower and the slope of the [$\alpha$/Fe]--[Fe/H] relation in it 
is flatter than those in the thick disk. This is because the star formation rates in
the subsystems differ.

Since the published results disagree, analyzing the relations between the 
relative abundances of $\alpha$-elements and metallicity and other parameters of
thin-diskstars based on a much larger statistical material seems very topical.
In this paper we analyze the chemical properties of thin-disk stars using
data from our compiled catalog of spectroscopically determined magnesium 
abundances (Borkova and Marsakov~2005). Magnesium is one of the best
studied $\alpha$-elements, because it has lines of various intensities and 
degrees of excitation in the visible spectral range in main-sequence F--G
stars. Almost all of the published magnesium abundances in dwarfs
and subgiants in the solar neighborhood determined by synthetic 
modeling of high-dispersion spectra as of December~2003 were gathered in 
the catalog. This catalog is several times larger than any homogeneous
sample that has been used until now to analyze the chemical evolution of 
the Galaxy.

The relative magnesium abundances in the catalog were derived from~1412 
spectroscopic determinations in 31~publications for 867~stars using
a three-pass iterative averaging procedure with a weight assigned to 
each primary source and each individual determination. The internal 
accuracy of the catalogued relative magnesium abundances for metal-rich 
($[Fe/H]>-1.0$~dex) stars is $\varepsilon$[Mg/Fe]$=\pm 0.05$~dex. The 
metallicities for the stars were obtained by averaging $\approx 2000$~[Fe/H]
determinations from 80~publications; the accuracy was estimated to be 
$\varepsilon$[Fe/H]$=\pm 0.07$~dex. The distances to the stars and their 
space velocities were calculated based on data from currently available 
high-precision catalogs. We used trigonometric parallaxes with errors smaller
than 25\,\% and, if these were lacking, photometric distances calculated 
from uvbyH$_\beta$ photometry. Based on a multicomponent model of the Galaxy
containing a disk,a bulge,and an extended massive halo from Allen and 
Santillan~(1991), we calculated the Galactic orbital elements by simulating
30~revolutions of the star around the Galactic center. The Galactocentric 
distance of the Sun was assumed to be 8.5~kpc, the rotational velocity of
the Galaxy at the solar Galactocentric distance was 220~km\,s$^{-1}$ and 
the velocity of the Sun with respect to the local standard of rest was 
(U$_\odot$, V$_\odot$, W$_\odot$)=($-11, 14, 7.5$)~km\,s$^{-1}$ (Ratnatunga 
et al.~1989). For 545~catalogued stars, we used the ages determined
from theoretical isochrones,trigonometric parallaxes,
and photometric metallicities from Nordstr$\ddot o$m et al.~(2004). For 
more details on stellar parameters, see Borkova and Marsakov~(2005) and 
Marsakov and Borkova~(2005).

\section*{IDENTIFICATION OF THIN-DISK STARS.}

Since our main goal is to analyze the relations between the chemical 
composition and other parameters of thin-disk stars, we identified the 
latter solely according to kinematic criteria. The thin-disk stars
are known to have low residual velocities with respect to the local 
standard of rest and nearly circular orbits with all of their points 
lying not high above the Galactic plane. Therefore, for reliability 
we identified this subsystem simultaneously by two kinematic
conditions. For nearby stars, a convenient criterion is
$V_{res}=\sqrt{U^2+V^2+W^2}<85$~km\,s$^{-1}$, where U, V, W, and V$_{res}$ 
are the space velocity components of the star and its residual velocity 
with respect to the local standard of rest, respectively. The other 
criterion does not depend on the location of the star in the Galactic
orbit and is based on the following formula suggested by Chiappini et 
al.~(1997): $\sqrt{Z^2_{max}+4\cdot e^2}<1.05$ where Z$_{max}$ is the 
maximum distance of the orbital points from the Galactic plane and $e$ is 
the orbital eccentricity of the star. Higher-velocity stars were included 
in the thick disk. We chose the criteria to minimize the number of 
stars with high and low relative magnesium abundances in the thin and thick
disks, respectively (for more details on the segregation of thick-disk
and thin-disk stars, see Marsakov and Borkova~(2005)). Figure~1a shows 
the distribution of stars in the V$_{res}$--$\sqrt{Z^2_{max}+4\cdot e^2}$
diagram, where the thin-disk and thick-disk stars are denoted by different
symbols. We additionally identi ed the stars for which the probabilities 
of belonging to the thin disk is a factor of 10~higher than the probability 
of belonging to the thick disk. The stars with the same high probabilities 
of belonging to the thickdiskwere selected and identified in the gure in a 
similar way. The technique for calculating the corresponding probabilities 
based on the dispersions of the space velocity components and the mean 
rotational velocity of the subsystem stars at the solar Galactocentric 
distance was taken from Bensby et al.~(2003). We see that the probability
criterion is more stringent and roughly corresponds to the criterion 
$\sqrt{Z^2_{max}+4\cdot e^2}<0.6$ for the thin disk. In this case, the 
first condition V$_{res} \leq 85$~km\,s$^{-1}$ is also satisfied automatically.
When more stringent criteria are used, a significant number of stars from
the original catalog turn out to be unidentified. To preserve the size of 
the sample used, below we will analyze the thin-disk properties using the 
stars identified according to our criteria, but we will always test
the reliability of the results using the stars selected according to 
this more stringent probability criterion.

Note that we cannot get rid of the overlap between the ranges of elemental 
abundances in the stars of the two disk subsystems by any combination of
any kinematic parameters.To illustrate this conclusion regarding the 
magnesium abundance, Fig.~1b shows the V$_{res}$--[Mg/Fe] diagram for the 
stars selected according to our criteria and Fig.~1c shows the same 
diagram for the stars selected according to the stringent probability 
criterion. The stars in which the magnesium abundances were taken from 
several sources and simultaneously with a total weight larger than unity 
were additionally identified in both diagrams. (Recall that the largest weight 
equal to unity in our catalog was assigned to the source (Edvardsson et al.~1993),
for whose stars the deviations of the derived magnesium abundances from their 
precalculated mean values were smallest.) We see from the diagrams that an 
appreciable percentage ($\approx 8$\,\%) of the stars with high relative 
magnesium abundances remain in the thin disk even for the most stringent
candidate selection and reliable [Mg/Fe] determination. After applying the 
stringent kinematic criterion $\approx 9$\,\% of the stars with low 
magnesium abundances also remain in the thick disk but the [Mg/Fe] ratios
were reliably determined in our catalog from several sources only for two 
of these stars. The overlap between the metallicity ranges of the subsystems 
is discussed in the next section.

\section*{THE METALLICITY -- MAGNESIUM ABUNDANCE RELATION.}

Figure~2 shows the [Fe/H]--[Mg/Fe] diagrams where the thin-disk and thick-disk 
stars are denoted by different symbols. (The stars of the so-called metal-poor 
tail of the thick disk i.\,e., those with [Fe/H]$<-1.25$ are not plotted in 
this gure.) A careful examination of the diagrams leads us to the following obvious 
conclusions. First, although the bulk of the thin-disk stars actually have 
$[Mg/Fe]<0.20$ and lie mostly in the range $-0.75 <[Fe/H ]<+0.35$ a sizeable 
fraction of them (52 of the 569 stars, i.\,e. $\approx 9$\,\%) have higher 
magnesium abundances at metallicities $-1.0 <[Fe/H]<- 0.30$. (Two stars 
(HD~83769 and BD~30$^\circ$ 18140) with thin-disk kinematics and with metallicities 
$[Fe/H ]<-2.0$ are disregarded in the subsequent analysis of the thin-disk 
properties.) In contrast, the bulk of the thick-disk stars lie in the metallicity 
range $-1.25 <[Fe/H ]<-0.3$ at $[Mg/Fe]>0.2$ but there are also stars with lower 
magnesium abundances in it (see also Marsakov and Borkova~2005); i.\,e.
not only the magnesium abundance ranges, but also the metallicity 
ranges overlap. Hence the position of the dip observed in the 
metallicity distributions of Galactic stars near $[Fe/H]\approx -0.5$ 
may be used as a criterion for separating the stars into the Galactic
subsystems only in the absence of space velocities for them (see, e.\,g.
Marsakov and Suchkov~1977). Another important result is that the 
metallicity--relative magnesium abundance relation in the thin disk differs in 
behavior from that in the thick disk although both subsystems exhibit a 
decrease in the relative magnesium abundance with increasing metallicity.
The median sequences for the thin and thick disks are plotted in Figs.~2b and 2c,
respectively. Since these cannot be constructed mathematically rigorously we 
drew them by eye halfway between the corresponding upper and lower envelopes 
in the diagrams. The circles mark the stars selected according to stringent 
probability criteria for their belonging to a particular subsystem and with 
weights of the final magnesium abundance larger than unity. In Fig.~2a,
both sequences are plotted simultaneously. We see that the sequence for 
the thin disk in the metallicity range ($-1.0 <[Fe/H]<-0.4$) lies 
systematically lower than that for the thick disk thereby confirming
the result of Bensby et al.~(2003). 

Let us now verify the result of 
Edvardsson et al.~(1993) using our larger sample of thin-disk stars
selected according to kinematic criteria, i.\,e. ascertain whether the 
positions of the thin-disk stars in [Mg/Fe] the [Fe/H]-–[Mg/Fe] diagram 
depend on their mean orbital radii? Figure~3 shows these diagrams for 
the thin-disk stars divided into four approximately equal (in number) 
R$_{m}$ ranges by the following values: 8, 8.5, and 9.1~kpc. Our median 
sequence is plotted in all diagrams. The circles denote the stars with
accurately determined magnesium abundances (more than two determinations)
and selected according to the probability criterion. We see from the 
figure that the behavior of the relation under study changes appreciably 
with increasing mean Galactocentric stellar orbital radius. Indeed, only 
the stars with the smallest mean orbital radii in Fig.~3a closely follow 
our median curve at $[Fe/H]<-0.4$~dex. However in the next diagram 
(Fig.~3b), the overwhelming majority of such metal-poor stars lie in the 
range $[Mg/Fe]<0.2$, with the number of points below the curve in the 
metallicity range ($-0.7\div -0.4$)~dex being considerably larger
than their number above the curve. At the largest distances, the 
[Mg/Fe]--[Fe/H] relation is almost linear (see the dashed line in Fig.~3d).
In this case, only a small number of stars with sharply enhanced magnesium 
abundances are observed above the curve. At the same time, it can be 
noticed that the metallicity range for the bulk of the stars with $[Mg/Fe]<0.2$ 
is displaced from ($0.5<[Fe/H]<+0.3$) for the nearest stars to ($-0.7<[Fe/H]<+0.2$)
for the farthest stars. This change in the behavior of the [Mg/Fe]--[Fe/H]
relation and the displacement of the metallicity range for stars with low 
relative magnesium abundances as their mean orbital radii increase confirm
the assumption by Edvardsson et al.~(1993) that the star formation rate in 
the thin disk decreases with Galactocentric distance.

Additional information can be obtained from the histograms in Fig.~4 which 
shows how some of the stellar population parameters change with Galactocentric 
distance. The distributions in the first three upper histograms and the second 
upper histogram of the second column are described by the sums of two Gaussians,
while the curves in the remaining histograms are eighth degree polynomials.
First, note that the distributions of all parameters for the stars with 
the smallest mean orbital radii drop out of the general trends for all 
of the remaining distances. Thus for example, in the first column the 
distribution of the nearest stars in relative magnesium abundance in Fig.~4a 
clearly shows a bimodal structure with the maxima near $[Mg/Fe]\approx 0.07$ and
$0.27$~dex (the latter relative magnesium abundance is typical of the
thick-disk stars). The histogram in the figure is well described by the 
sum of two Gaussians whose parameters were determined by the maximum 
likelihood method. In this case,the probability of erroneously rejecting 
the hypothesis about the description of the distribution by one Gaussian 
against its alternative representation by the sum of two Gaussians is 
P$_N \approx 1$\,\% (see Marsakov et al.~(1984) for more details on determining 
the statistical signi cance of tting the distribution by the sum of 
Gaussians and Martin~(1971) for a justification of the method). Edvardsson et 
al.~(1993) also pointed out that the magnesium abundance distribution may 
be bimodal for the stars with mean orbital radii smaller than the solar 
Galactocentric distance. At larger distances, the overwhelming majority of 
stars lie at $[Mg/Fe]<0.2$ (see the first column in Fig.~4). All of them 
have unimodal distributions, but there is a clear tendency for the positions
of the distribution maxima to be displaced toward the higher relative 
magnesium abundances with increasing mean stellar orbital radii.

The metallicity distributions of stars in the second column behave even 
more expressively. Here, the stars closest to the center in Fig.~4b also 
exhibit a bimodal distribution (P$_N \approx 4$\,\%). The primary (large)
maximum lies near $[Fe/H]\approx -0.30$~dex, while the secondary maximum 
is near $\approx +0.05$~dex. For the farther stars in Fig.~4f (P$_N < 1$\,\%),
the primary maximum is suddenly displaced to the right flank of the diagram 
into the region of positive metallicities. In other words, stars with 
solar heavy-element abundances dominate among the stars close to the Sun
with such an orbital characteristic. As the mean orbital radii increase 
further the distributions show a systematic increase in the relative 
number of stars in the metal-poor group. Although the last two 
distributions cannot be described by the sums of two Gaussians, the 
polynomial clearly shows that the primary (and already unique) maximum 
for the farthest stars (Fig.~4n) is again near $[Fe/H]\approx -0.30$ while
the percentage of stars with solar metallicity becomes low. This behavior 
manifests itself in the existence of a negative radial metallicity 
gradient that results (in the opinion of Edvardsson et al.~(1993)) from 
a decrease in the star formation rate with increasing Galactocentric 
distance.

Clearly these distributions of elemental abundances (differing sharply 
from others) in the stars whose orbits lie almost entirely within the 
solar circle are attributable to their higher (on average) eccentricities.
The distributions of the latter are given in the third column. We see 
that not only low eccentricity stars are absent in the histogram of
the nearest stars (Fig.~4c), but also it is bimodal. (Although the 
maximum likelihood method here yields a high probability of erroneously 
rejecting the hypothesis about the description of the distribution by
one Gaussian against its alternative representation by the sum of two 
Gaussians (P$_N >5$\,\%), a clearly distinguishable dip is observed in 
the histogram.) Since the lower-eccentricity stars from inner Galactic
regions do not reach the solar orbital radius, the left hump in the 
histogram is truncated. However the dip in the distribution is 
distinguished quite clearly; i.\,e., the sample of stars with small 
$R_m$ seems to be kinematically inhomogeneous -- it most likely contains
stars of two different thin-disk populations. While remaining unimodal,
the histograms of stars with orbital radii larger than the solar one 
show a gradual change in the position of the distribution maximum
with increasing $R_m$ toward the higher eccentricities; i.e., a 
transition from the dominance of one population to that of the other 
is observed. (The position of the maximum at $e<0.1$ here is explained 
by the absence of selection by eccentricity at fairly large $R_m$.) The 
age distributions are also consistent with the assumption that the 
thin disk has a multicomponent structure: although all of them are 
unimodal, the dispersion for the stars closest to the Galactic center 
in Fig.~4d is largest. In this case, as we see from the histograms
in the fourth column the position of the distribution maximum is 
displaced appreciably toward the older ages with increasing $R_m$ 
beginning from the solar orbital radius.

Let us examine whether the situation will change if the apogalactic 
orbital radii of the stars are considered as their presumed birthplaces.
Note that the mean orbital radii are preferably used in this capacity 
only because they are less subject to the distorting effects of external 
gravitational perturbations (i.\,e. relaxation). However under the 
assumption that the orbits of the stars do not undergo significant 
changes since their birth in the overwhelming majority of cases, it 
would be more appropriate to use their maximum Galactocentric orbital 
radii (for an insignificant role of relaxation in the thin disk, see 
Marsakov and Shevelev~(1994)). Figure~5 shows the [Fe/H]--[Mg/Fe] 
diagrams for the stars divided into four approximately equal (in number)
groups in Ra by the values of 8.8, 9.4, and 10.3~kpc. These diagrams 
also show a gradual decrease in the relative number of lower metallicity 
$([Fe/H ]<0.4)$ stars above the median curve and a displacement of the 
metallicity range for magnesium-poor stars with increasing orbital 
radius. The changes here are not so sharp as those for the mean 
orbital radii, but, as a result, the overwhelming majority of stars 
with the largest apogalactic orbital radii also lie along the straight 
line that starts from $[Fe/H]\approx -0.7$ and passes below the median 
curve up to $[Fe/H]\approx -0.3$ (see the dashed line in Fig.~5d).
Thus assuming the maximum Galactocentric distances of the stars to 
be their birthplaces, we also conclude that the increase in the star 
formation rate with decreasing Galactocentric distance is responsible 
for the existence of a negative radial metallicity gradient.

The distribution of stars with the smallest apogalactic orbital 
radii in relative magnesium abundance in Fig.~6a cannot be described 
by the sum of two Gaussians, as in Fig.~4a, but most of the stars
with $[Mg/Fe]>0.2$ fell precisely into this group and they formed a 
clearly distinguishable structure in the histogram. As in the case 
of the division into $R_m$ ranges, the remaining histograms in the 
first column also exhibit a small systematic displacement of the 
positions of the distribution maxima toward the higher relative 
magnesium abundances in the disk stars with increasing apogalactic 
orbital radius. However this displacement is too small to be able to
give rise a significant radial gradient in relative magnesium 
abundance. The [Fe/H] distribution of stars close to the Galactic 
center in Fig.~6b is described by the sum of two Gaussians at a 
high significance level (P$_N <1$\,\%); the positions of the maxima 
of these Gaussians coincide, within the error limits, with those 
found in Fig.~4a. We see from the figure that the numbers of stars 
under the two Gaussians initially become equal as the apogalactic 
orbital radii increase further (Fig.~6f) and, subsequently as $R_a$
increases, the relative number of stars in the metal-poor group 
becomes so large that the metal-rich hump in the histogram 
disappears, showing only a small excess in the distribution (Fig.~6n).
(The last two distributions as well as those in Figs.~4a and 4n cannot 
be described by the sum of two Gaussians at a statistically signi 
cant level, but polynomials allow the displacement of the positions 
of the distribution maxima with increasing stellar orbital radii to 
be traced.) This behavior of the metallicity distributions for 
stars born at different Galactocentric distances, first, gives rise 
to a radial metallicity gradient and second, points to the possible 
existence of two stellar populations in the disk. A distinct bimodal 
eccentricity distribution of stars with the smallest orbital radii
(the significance of its description by the sum of two Gaussians is 
P$_N\ll 1$\,\%) and a systematic displacement of the positions of 
the eccentricity distribution maxima with increasing orbital radii 
can be seen from this gure even more clearly than from Fig.~4.
(However this displacement can be entirely explained by the analytical 
dependence of the eccentricities on orbital radii.) The age histograms 
for the stars of the samples in $R_a$ also exhibit a displacement of
the distribution maxima toward the older ages with increasing 
Galactocentric distance, but this is not so clear as that in Fig.~4.

\section*{RELATION BETWEEN THE CHEMICAL COMPOSITION AND OTHER PARAMETERS OF STARS}

The thin disk was formed over a considerably longer period than 
all of the older subsystems. Therefore it would be natural to 
expect a correlation of the chemical composition of stars in 
this subsystem with their ages and orbital elements. The 
existence of vertical and radial metallicity gradients in the 
Galactic thin disk is believed to have been firmly established; 
nevertheless, let us refine these gradients using the data of 
our sample and simultaneously determine the same gradients
in relative magnesium abundances. The diagrams of interest 
are shown in Fig.~7. The circles mark the stars selected 
according to the probability criterion and with accurately 
determined magnesium abundances. The straight lines in the 
diagrams were drawn by the least-squares method and their slopes 
define the corresponding gradients. All three metallicity gradients 
coincided within the error limits, with the typically derived 
gradients for mixed-age Galactic diskstars (see, e.\,g. Shevelev 
and Marsakov~(1995) and references therein): 
grad$_Z$[Fe/H]$=-0.29 \pm 0.06$~kpc$^{-1}$,
grad$_{Ra}$[Fe/H]$= -0.05 \pm 0.01$~kpc$^{-1}$
and grad$_{Rm}$[Fe/H]$= -0.04 \pm 0.01$~kpc$^{-1}$ with the
correlation coefficients $r =0.21 \pm 0.05$, $0.25 \pm 0.04$
and $0.15 \pm 0.04$ respectively. The large size of the original 
sample of thin-disk stars ensured a high reliability of the 
results obtained; in addition our test showed that all of the 
gradients calculated from the strictly selected stars (open 
circles) coincided with the above ones, within the error limits. 
The diagrams to determine the gradients in relative magnesium 
abundance are shown in the right column of the gure. The vertical 
gradient was found to be nonzero far beyond the error limits: 
grad$_Z$[Mg/Fe]$=(0.13 \pm 0.02)$~kpc$^{-1}$ at 
$r =0.28 \pm 0.04$ (Fig.~7b). The result is stable and the 
gradient derived from the strictly selected stars is found to 
be almost the same ($0.15 \pm 0.05$~kpc at $r =0.28$). (Note that 
the farthest points in the diagrams that often determine the 
correlation are automatically discarded in this case. This is 
further evidence that the result is reliable.) In contrast, the 
radial gradient in magnesium abundance turned out to be zero 
outside the error limits under both assumptions about the
birthplaces of the stars (see Figs.~7d and 7f).

Twarog~(1980) was the first to derive the age--metallicity relation 
in the Galactic disk from F2--G2 and argued that it was unambiguous. 
However it was subsequently proven that the relation was by no 
means unambiguous and there was a significant spread in metallicity 
among the stars of any ages (see Marsakov et al.~1990; Feldzing et 
al.~2001), which gave reason to suggest that there was no 
age--metallicity relation in the disk.The t--[Fe/H] diagram for 
the stars of our sample is shown in Fig.~8a. Note that the 
absolute errors in the ages we use, which were determined by 
Nordstr$\ddot e$m et al.~(2004), are less than 2~Gyr for $\sim 55$\,\%
of the stars and are larger for the remaining stars, reaching 
$\approx 7$~Gyr for some of the stars. As our test showed, they are 
large mostly for old stars; therefore, we analyze here the 
behavior of only stars with $t<10$~Gyr.

The correlation coefficient calculated only from the stars selected 
according to stringent criteria at $t<10$~Gyr (Fig.~8a) is larger 
than zero outside $3\sigma$: $r = 0.30 \pm 0.07$. The probability 
that the determinations of the gradient from the same number 
of two uncorrelated quantities will yield no lower correlation
coefficient is P$_N \approx 1$\,\%. We see from the figure that the 
correlation is attributable mainly to the absence of young metal-poor 
stars in the thin disk: the lower left corner in the diagram is 
empty (the dashed inclined line in Fig.~8a). Such an effect is 
revealed by absolutely all previous studies and is not the result 
of selection (for more detail, see Shevelev and Marsakov~1993). 
An event that led to a continuous increase in the mean heavy-element 
abundances and to a decrease in the metallicity spread in the younger
generations of stars probably occurred $\sim 4$~Gyr ago. If the 
sample of thin-disk stars is limited to an age $t>4$~Gyr then the 
age--metallicity correlation will actually disappear completely. The 
t--[Mg/Fe] diagram in Fig.~8b shows the same large correlation coefficient:
$r =0.27 \pm 0.08$ at P$_N \approx 1$\,\%. As a result, it turns out 
that the relative magnesium abundance in the thin-disk stars slightly 
increases with age. The correlation here is attributable not only to 
the sudden increase in the [Mg/Fe] spread after $\approx 9$~Gyr since,
if the sample is limited to this age, then both the slope and the
correlation coefficient will be almost constant. Here, when the sample 
is limited solely to old stars the correlation does not disappear 
completely although not only the correlation coefficient, but also the 
slope of the regression line decrease sharply remaining nonzero 
outside the error limits. The significance of the existence of both 
correlations is confirmed by the behavior of the strictly selected 
thin-disk stars with accurately determined magnesium abundances
(open circles in the diagrams),according to which the values found 
remain constant.

The bimodal distributions of thin-disk stars in iron and magnesium 
abundances and in Galactic orbital eccentricity (see Figs.~4 and 6) 
and the existence of an abrupt change in metallicity with age (Fig.~8)
most likely suggest that the population in the subsystem is 
inhomogeneous. Let us check whether the gradients in chemical 
composition in the disk populations of different ages also differ. To 
this end, we divide all of the thin-disk stars into two groups by
$t =4$~Gyr. As we see from the upper two diagrams in Fig.~9 the 
vertical metallicity gradient for the young group is considerably 
steeper than that for the old group: grad$_Z$[Fe/H]$=-0.58 \pm 0.13$~kpc$^{-1}$
and $-0.16 \pm 0.07$~kpc$^{-1}$ with the correlation coefficients 
$r = 0.36 \pm 0.04$ and $0.13 \pm 0.05$ and the probabilities of 
random occurrence of the correlations $\ll 1$\,\% and $\approx 2$\,\% 
for the young and old groups, respectively. The derived gradients 
change only slightly if the sample is limited to the stars selected 
according to stringent criteria and if the five farthest points are
eliminated from each diagram. That the gradient in the young group 
proved to be steeper closely agrees with the result obtained by 
Shevelev and Marsakov~(1995) from a considerably larger sample, but 
with photometric metallicities and distances. A decrease in the 
vertical metallicity gradient with increasing stellar age can be 
observed in the case of a continuous increase in the stellar 
residual velocity dispersion with time, i.\,e., relaxation. Indeed, 
the distortion of the stellar orbits with time must cause the 
gradient to be blurred. However Fig.~9b clearly shows that the 
shallow gradient in the old group is attributable not to an 
increase in the maximum distance from the Galactic plane for all 
old stars of any metallicity but, on the contrary to the avoidance 
of large $Z_{max}$ by the metal-poor stars (the empty lower right 
corner in the diagram). As a result, the lower envelope of the 
diagram even shows a tendency for the mean metallicity to increase 
with age. Therefore, we are inclined to believe that relaxation 
in the thin disk plays no crucial role. (Marsakov and Shevelev~(1994) 
reached the same conclusion by analyzing the age dependences of the 
parameters of the velocity ellipsoids for thin-disk F stars.)

As we see from the lower diagrams in Figs.~9c and 9d, the vertical 
gradient in relative magnesium abundance for the young group was also 
found to be slightly steeper than that for the old group: 
grad$_Z$[Mg/Fe]$=0.12 \pm 0.04$~kpc$^{-1}$ and $0.07 \pm 0.02$~kpc$^{-1}$ 
at $r =0.23 \pm 0.08$ and $0.18 \pm 0.06$ for the young and old group,
respectively. In both cases, despite the small correlation coefficients,
the vertical gradients in relative magnesium abundance were found 
to be nonzero outside $3\sigma$ for the probabilities of their 
random occurrence P$_N <1$\,\%. Remarkably Fig.~9d reveals a paradox 
in the older group similar to that in Fig.~9b: the upper right corner 
in the diagram unexpectedly proved to be empty; i.\,e., while the mean 
relative abundance shows a general tendency to increase with $Z_{max}$,
a deficit of magnesium-rich ($[Mg/Fe ]>0.25$~dex) stars far from the 
Galactic plane ($Z_{max} >0.3$~kpc) is observed in the thin disk.
Note also that most of the stars in the young group have metallicities 
$[Fe/H]> 0.4$~dex and magnesium abundances $[Mg/Fe]<0.15$~dex and lie 
above the Galactic plane mainly at $Z<400$~pc while a significant 
fraction of the stars in the old group exceed considerably these 
limits in all three parameters.

As we see from the diagrams in Fig.~10, all of the radial metallicity 
gradients are nonzero far beyond the error limits with P$_N \ll 1$\,\%
(see the inscriptions in the corresponding panels). Neither the removal
of far points in the diagrams nor the use of solely strictly selected 
stars change the results. As a result, we see that the radial 
metallicity gradients show an indistinct tendency for the absolute 
value to decrease with increasing age (the differences in gradients lie
outside $2\sigma$ in $R_m$ and do not exceed the errors in these 
regression coefficients in $R_a$). The same small rise in the radial 
metallicity gradient in the thin disk with age was also found by 
Shevelev and Marsakov~(1995).

There are no radial gradients in relative magnesium abundance in both 
age groups, as in the case of the complete original sample of mixed-age 
stars (see Fig.~7).

\section*{THE FORMATION OF THE THIN DISK}

Analysis of the relations between the relative magnesium abundance 
and other parameters of the stellar populations makes it possible to 
better understand the formation history of the Galactic subsystems. 
Indeed, even the difference in the mean relative magnesium abundances 
between the thick-disk and thin-disk stars suggests that the bulk of 
the thick-disk stars formed long before the onset of massive star
formation in the thin disk. However the overlap between the ranges 
in both metallicity and magnesium abundance for these subsystems is 
also observed. Hence, the star formation process most likely did not
cease completely in the entire Galaxy. The [Mg/Fe] ratio begins to 
decrease with increasing metallicity in the thin disk immediately 
after the formation of the first stars in it, i.\,e., from 
$[Fe/H]\approx -1.0$~dex. This is considerably farther to the left 
in the diagram than in the thick disk where the point of a sharp
decrease is observed at $[Fe/H]\approx 0.5$~dex. Hence, in the 
metal-poor interstellar matter from which the first thin-disk stars 
subsequently began to form, the enrichment with SNe\,II ejecta was 
less intense before this. It is quite probable that such metal-poor 
matter with a high relative magnesium abundance came into the thin 
disk as a result of accretion from regions with a different history 
of chemical evolution. The fast decrease in the relative magnesium 
abundance in the thin disk as the metallicity increases from 
$\approx -1.0$~dex to $\approx -0.7$~dex suggests that the star 
formation rate in it was initially low but it then suddenly increased,
which subsequently led to a attening of the [Fe/H]--[Mg/Fe] relation.
This event occurred $\sim 9$~Gyr ago, which is also evidenced by the 
sharp increase in the number of thin-disk stars younger than this age.
Subsequently when passing to stars with metallicities higher than 
the solar value, the slope of the [Mg/Fe]--[Fe/H] relation virtually 
vanishes, which is indicative of a new increase in the star formation
rate $\sim 4$~Gyr ago and stabilization of the ratio of the 
contributions from supernovae (SN\,II/SN\,Ia) to the enrichment of 
the interstellar medium in the thin disk since then. In contrast, in 
the thick disk star formation virtually ceased altogether (for some
time?) after the onset of massive SN\,Ia explosions, as suggested by 
the observed steep decrease in [Mg/Fe] near [Fe/H]$\approx -0.5$~dex.

The change in the behavior of the [Mg/Fe]--[Fe/H] relation with 
stellar orbital radius shows that the star formation rate closer to 
the Galactic center is higher than that on the periphery. Moreover it
seems that star formation within the solar circle of the Galaxy has 
never been interrupted, but only slowed down before the massive 
formation of thin-disk stars. In contrast, the first thin-disk stars at
great Galactocentric distances appeared only after the long phase of 
star formation delay. This follows from the presence of a distinct 
jump in the [Mg/Fe] ratio at metallicities $[Fe/H]<-0.4$~dex for stars
with large orbital radii. The larger [Mg/Fe] ratios at high 
metallicities in the stars within the solar circle suggest that the 
star formation rate remains there higher even at present. The clear 
deficit of stars with metallicities higher than the solar value there 
is also indicative of a lower star formation rate at great 
Galactocentric distances. Such differing histories of chemical 
evolution along the radius led to the formation of a radial metallicity 
gradient in the Galactic disk. However the corresponding gradient
in relative magnesium abundance cannot be found, most likely because 
it is insignificant compared to the spread in [Mg/Fe] in the subsystem.
The existence of nonzero vertical gradients in both metallicity and
relative magnesium abundance far beyond the error limits in the 
thin disk probably suggests that the fall of a metal-poor gas to the 
disk, the star formation, and the contraction of the interstellar 
medium in the forming subsystem are simultaneous processes.

The time-varying star formation rate in the thin disk could not but 
give rise to the age--metallicity and age--magnesium abundance 
relations. Since the chemical and spatial--kinematic properties of 
stars of different ages turned out to be different, the thin-disk 
structure is probably not homogeneous. This is also suggested by 
the bimodality of the metallicity function in it and by the 
difference in the kinematic parameters of the groups of stars with 
different metallicities (see Marsakov and Suchkov~1980). However
reliable segregation of stars into populations inside it does not 
seem possible. This is because the stars that are currently in 
the solar neighborhood arrive here from regions with different 
(as we now see) star formation histories and the stars of different 
ages born far from one another may prove to have the same chemical 
composition.

Despite the long history of studying this question an adequate 
model for the formation of the Galactic thin disk and its evolution 
cannot be constructed so far. A thorough analysis of the detailed 
chemical composition of a large number of the nearest stars can 
help obtain additional information about these processes.

{\bf Acknowledgements}: 
We are grateful to the anonymous referee for valuable remarks 
that forces us to present our results in a more reasoned way.
This study was supported by Russian Federal Agency on Education 
(projects RNP 2.1.1.3483 and RNP 2.2.3.1.3950) and grant of ROSNAURA 
02.438.11.7001

\newpage

\begin{enumerate}
\item[Fig.1. - ]{Relation between the residual velocities with respect to the 
local standard of rest and $\sqrt{Z^2_{max}+4\cdot e^2}$ (a), the [Mg/Fe]
ratios for all of the sample stars separated according to our criteria into 
the two disk subsystems (b) and only for the stars selected in the corresponding 
subsystems according to stringent probability criteria (c): the pluses and 
triangles indicate the thin-disk and thick-disk stars, respectively; the circles 
markthe stars selected according to stringent criteria (a) and with
accurately determined relative magnesium abundances (b,c).}
    
\item[Fig.2. - ]{Relation between metallicity and relative magnesium abundance (a)
for the stars of both disk subsystems (b) for the thin-disk stars, and (c)
for the thick-disk stars: the crosses and triangles indicate the thin-disk
and thick-disk stars, respectively. The circles mark the stars selected 
according to the probability criterion and with accurately determined
magnesium abundances. The broken curves represent the median lines of the 
relations for the thin (a) and thick (b) disks drawn by eye halfway between 
the upper and lower envelopes. The error bars are shown.}
    
\item[Fig.3. - ]{Relation between metallicity and relative magnesium 
abundance for the thin-disk stars separated into four ranges in
mean orbital radii by R$_m$= 8, 8.5 and 9.1~kpc. The circles 
mark the stars selected according to the probability criterion 
and with accurately determined magnesium abundances. The broken 
curves in all panels represent the median line for the thin disk.
The error bars are shown.}   

\item[Fig.4. - ]{Distributions for thin-disk stars separated into four 
ranges in $R_m$ (first column) of relative magnesium abundance, 
(second column) of metallicity, (third column) of galactic orbit 
eccentricities, and (fourth column) of ages. Distributions on the 
first three panels on the first line and on the second panel of the 
second line approximated by the sums of two Gauss functions and by 
the eight power polynomials on the rest panels.}   

\item[Fig.5. - ]{Same as Fig.~3 for the thin-diskstars divided into four 
ranges in apogalactic orbital radii by $R_a$= 8.8, 9.4 and 10.3~kpc.}   

\item[Fig.6. - ]{Same as Fig.~4 for the thin-disk stars divided into 
four ranges in $R_a$}   

\item[Fig.7. - ]{Relations of metallicity and relative magnesium abundance 
in the thin-disk stars to the maximum distance of their orbital 
points from the Galactic plane (a,b),apogalactic distance (c,d), 
and mean orbital radius (e,f). The circles markthe stars with 
accurate [Mg/Fe] ratios selected in the thin disk according to 
a strict probability criterion. The solid lines represent the 
regression lines. The error bars along with the gradients and 
the correlation coefficients are indicated.}   

\item[Fig.8. - ]{Relations between the age of thin-disk stars and metallicity (a)
and relative magnesium abundance (b). The notation is the same as that
in Fig.~7.}   

\item[Fig.9. - ]{Relations between the maximum distance of thin-disk stars 
from the Galactic plane and their metallicity (a,b) and relative 
magnesium abundance (c,d): for the stars younger and older than 4~Gyr 
on the left and on the right, respectively.}   

\item[Fig.10. - ]{Relations between metallicity and mean stellar orbital radius 
(a,b)and apogalactic radius (c,d): for the stars younger and older 
than $4$~Gyr on the left and on the right respectively. The 
notation is the same as that in Fig.~7.}   

\end{enumerate}

\end{document}